# THE BINARY GALAXIES NGC 147 AND NGC 185

SIDNEY VAN DEN BERGH

Dominion Astrophysical Observatory, Herzberg Institute of Astrophysics,

National Research Council of Canada, 5071 West Saanich Road, Victoria, British Columbia,

V8X 4M6, Canada; vdb@dao.nrc.ca

## ABSTRACT

Contrary to a previously published claim it is found that the spheroidal galaxies NGC 147 and NGC 185 probably form a stable binary system. Distance estimates place this pair on the near side of the Andromeda subgroup of the Local Group. The fact that this system has probably remained stable over a Hubble time suggests that it does not have a plunging orbit that brings it very close to M 31. It is noted that the only two Local Group galaxy pairs, in which the components have comparable masses, also have similar morphological types. NGC 147 and NGC 185 are both spheroidals, while the LMC and SMC are both irregulars. This suggests that protogalaxies of similar mass that are spawned in similar environments evolve into objects having similar morphologies.

## 1. INTRODUCTION

The low-luminosity galaxies NGC 147 and NGC 185 have played an important role in the development of modern ideas on stellar populations. They were first studied in detail by Baade (1944). Subsequently Baade (1963) wrote "In August and September [of 1943] I resolved the central part of M 31 and the two nearby companions in rapid succession. Mayall had told me that NGC 185 (which forms a pair with NGC 147, some 12° from M 31) has a radial velocity very similar to that of M 31, and that he suspected on this basis that they might be nearby and associated with the Andromeda Nebula system. So I tried NGC 185 and NGC 147, and indeed they were easy to resolve". Baade (1951) pointed out that NGC 147 is an example of a dust-free galaxy containing only stars of Population II. On the other hand NGC 185 contains prominent dust patches and a sprinkling of bright young blue stars. This led Baade to conclude that "these examples show in a striking manner the intimate relationship between the presence of dust and the appearance of type I stars. No dust, no [P]opulation I, particularly no B and O stars."

## 2. ARE NGC 147 AND NGC 185 GRAVITATIONALLY BOUND?

Since it is usually easier to determine radial velocities from emission lines than from stellar absorption features, Ford, Jacoby & Jenner (1977) determined the radial velocities of NGC 147 and NGC 185 by observing a few planetary nebulae that they had discovered in these galaxies. From observations of the planetary nebulae NGC 147-1 and NGC 185-1 these authors concluded that the radial velocities of these galaxies differed by 40 km s$^{-1}$. They then noted that this value was incompatible with the requirement that these two galaxies are bound at the 3$\sigma$ level. "Thus, we conclude that NGC 185 and NGC 147 are not a gravitationally bound system." This conclusion, which is based on observations of only one planetary in each galaxy, is suspect because the internal velocity dispersion in each of these two galaxies is ~(23 ± 5) km s$^{-1}$ (Bender, Paquet & Nieto 1991). Using the digital spectrum of a relatively metal-poor G8 star as a template Bender et al. find radial velocities of (-193 ± 3) km s$^{-1}$ and (-202 ± 3) km s$^{-1}$ for NGC 147 and NGC 185, respectively. The resulting radial velocity difference is $\Delta V = (9.0 \pm 4.2)$ km s$^{-1}$.

For two galaxies to be gravitationally bound (Davis et al. 1995, Barmby & Huchra 1998) they must obey the relation

$$\Delta V^2 R_p / 2GM \leq \sin^2 \alpha \cos\alpha < 0.385, \qquad (1)$$

in which $\Delta V$ is the difference in their observed radial velocities, $R_p$ is their projected separation, M is the system mass and $\alpha$ is the angle between the line connecting the two galaxies and the plane of the sky. From a rediscussion of recent distance determinations to NGC 147 (Mould, Kristian & Da Costa 1983, Saha, Hoessel & Mossman 1990, Davidge 1994, Han et al. 1997, Salaris & Cassisi 1998) and to NGC 185 (Saha & Hoessel 1990, Lee, Freedman & Madore 1993, Martínez-Delgado & Aparicio 1998) van den Bergh (2000) concludes that the distance to this pair is (660 ± 30) kpc. Substituting this value into Eqn. (1) yields a combined mass M ≥ 2.7 x 10$^8$ M$\odot$, if these two objects are to be gravitationally bound. From this value, and luminosities $M_B$ = -14.4 for NGC 147 and $M_B$ = -14.8 for NGC 185, one finds that $M/L_B \geq 1.2$ (in solar units) for these two galaxies to constitute a physical pair. By combining spectroscopically determined internal velocity dispersion measurements with surface photometry Bender, Paquet & Nieto (1991) find $M/L_B = 7 \pm 3$ in (solar units) for NGC 147 and $M/L_B = 5 \pm 2$ for NGC 185. These values comfortably exceed the minimum value $M/L_B = 1.2$ required for NGC 147 and NGC 185 to be gravitationally bound. Additional arguments in favor of the hypothesis that NGC 147 and NGC 185 form a physical pair are that (1) the mean apparent magnitudes of the RR Lyrae variables on the Gunn system are <g> = 25.25 ± 0.10 for NGC 147 (Saha, Hoessel & Mossman (1990) and <g> = 25.20 ± 0.05 for NGC 185 (Saha & Hoessel (1990). This shows that these two galaxies, which have similar reddening values of E(B-V) = 0.17 for NGC 147 (Mould et al. 1983) and E(B-V) = 0.19 ± 0.03 for NGC 185 (Lee, Freedman & Madore 1993), are (within the uncertainties of the measurements) at the same distance. Furthermore (2) NGC 147 and NGC 185 are separated by only 0.972 on the sky. At a distance of 660 kpc this corresponds to a (projected) linear separation of only 11 kpc.

### 3.    CONCLUSIONS

All of the presently available data are consistent with the conclusion that NGC 147 and NGC 185 form a gravitationally bound system. This would make NGC 147 + NGC 185 and the LMC + SMC the only Local group binary systems with components that have roughly equal mass (luminosity). It may be no coincidence that both NGC 147 and NGC 185 are spheroidal galaxies, and that the Large Cloud and the Small Cloud are both Magellanic irregulars. In other words there may be a tendency for proto-galaxies of similar mass, that form in the same environment, to develop into objects that have similar morphologies. It is, however, puzzling that NGC 147 is presently dust and gas free (Young & Lo 1997, Sage, Welch & Mitchell 1998), whereas NGC 185 contains gas dust and a smattering of young stars (Baade 1951).

If the distance to the NGC 147 + NGC 185 pair is (660 ± 30) kpc then these objects lie closer to us than the Andromeda galaxy, for which Freedman & Madore (1990) obtained a distance of 770 ± 42 kpc. From the presence of RR Lyrae stars in both NGC 147 (Saha et al.1990) and in NGC 185 (Saha & Hoessel 1990) it follows that each of these galaxies was formed ≳ 10 Gyr, i.e. the members of this pair have each existed for a Hubble time. If the binary system itself has survived for such a long time then it cannot be on a plunging orbit within the Andromeda subgroup of the Local Group. This is so because a close encounter with M 31 (or its satellites M 32 and NGC 205) would probably have disrupted the NGC 147 + NGC 185 binary system.

# REFERENCES

Baade, W. 1944, ApJ, 100, 147

Baade, W. 1951, Publ. Obs. U. Michigan, Vol. X, p. 7

Baade, W. 1963, Evolution of Stars and Galaxies, Payne-Gaposchkin, C.

(Cambridge: Harvard University Press), p. 51

Barmby, P. & Huchra, J.P. 1998, AJ, in press, astro-ph/9709270

Bender, R., Paquet, A. & Nieto, J.-L. 1991, A&A, 246, 349

Davidge, T.J. 1994, AJ, 108, 2123

Davis, D.S., Bird, C.M., Mushotzky, R.F. & Odewahn, S.C. 1995, ApJ, 440, 48

Ford, H.C., Jacoby, G. & Jenner, D.C. 1977, ApJ, 213, 18

Freedman, W.L. & Madore, B.F. 1990, ApJ, 365, 186

Han, M. et al. 1997, AJ, 113, 1001

Lee, M.G., Freedman, W.L. & Madore, B.F. 1993, AJ, 106, 964

Martínez-Delgado, D. & Aparicio, A. 1998, AJ, 115, 1462

Mould, J., Kristian, J. & Da Costa, G.S. 1983, ApJ, 270, 47

Sage, L.J., Welch, G.A. & Mitchell, G.F. 1998, ApJ (in press)

Saha, A. & Hoessel, J.G. 1990, AJ, 99, 97

Saha, A., Hoessel, J.G. & Mossman, A.E. 1990, AJ, 100, 108

Salaris, M. & Cassisi, S. 1998, MNRAS (in press)

van den Bergh, S. 2000, The Galaxies of The Local Group (Cambridge:

Cambridge University Press)

Young, L.M. & Lo, K.Y. 1997, ApJ, 476, 127